\documentstyle[12pt]{article}

\newcounter{sectionc}\newcounter{subsectionc}\newcounter{subsubsectionc}
\renewcommand{\section}[1] {\vspace*{0.6cm}\addtocounter{sectionc}{1}
\setcounter{subsectionc}{0}\setcounter{subsubsectionc}{0}\noindent
        {\normalsize\bf\thesectionc. #1}\par\vspace*{0.4cm}}
\renewcommand{\subsection}[1] {\vspace*{0.6cm}\addtocounter{subsectionc}{1}
        \setcounter{subsubsectionc}{0}\noindent
        {\normalsize\it\thesectionc.\thesubsectionc. #1}\par\vspace*{0.4cm}}
\renewcommand{\subsubsection}[1]
{\vspace*{0.6cm}\addtocounter{subsubsectionc}{1}
        \noindent
{\normalsize\rm\thesectionc.\thesubsectionc.\thesubsubsectionc.
        #1}\par\vspace*{0.4cm}}

\newcounter{appendixc}
\newcounter{subappendixc}[appendixc]
\newcounter{subsubappendixc}[subappendixc]

\renewcommand{\appendix}[1] {\vspace*{0.6cm}
        \refstepcounter{appendixc}
        \setcounter{figure}{0}
        \setcounter{table}{0}
        \setcounter{equation}{0}
        \renewcommand{\thefigure}{\Alph{appendixc}.\arabic{figure}}
        \renewcommand{\thetable}{\Alph{appendixc}.\arabic{table}}
        \renewcommand{\theappendixc}{\Alph{appendixc}}
        \renewcommand{\theequation}{\Alph{appendixc}.\arabic{equation}}
        \noindent{\bf Appendix \theappendixc #1}\par\vspace*{0.4cm}}

\newcounter{itemlistc}
\newcounter{romanlistc}
\newcounter{alphlistc}
\newcounter{arabiclistc}

\begin{document}
\baselineskip=8truemm

\centerline{\normalsize\bf NEW PERSPECTIVES IN COMPLEX GENERAL
RELATIVITY}
\vspace{0.6cm}
\centerline{\footnotesize GIAMPIERO ESPOSITO}
\baselineskip=13pt
\centerline{\footnotesize\it Istituto Nazionale di Fisica Nucleare,
Sezione di Napoli}
\baselineskip=12pt
\centerline{\footnotesize\it Mostra d'Oltremare Padiglione 20,
80125 Napoli, Italy}
\baselineskip=12pt
\centerline{\footnotesize\it Dipartimento di Scienze
Fisiche, Mostra d'Oltremare Padiglione 19, 80125 Napoli, Italy}
\centerline{\footnotesize E-mail: esposito@napoli.infn.it}
\vskip 1cm
\baselineskip=13pt

\leftskip=12mm
\rightskip=12mm

{\small
\noindent
\hskip 12mm {\bf Abstract.} In complex general relativity, Lorentzian
space-time is replaced by a four-complex-dimensional
complex-Riemannian manifold, with holomorphic connection and
holomorphic curvature tensor. A multisymplectic analysis shows
that the Hamiltonian constraint is
replaced by a geometric structure linear in the holomorphic
multimomenta, providing some boundary conditions are imposed
on two-complex-dimensional surfaces. On studying such boundary
conditions, a link with the Penrose twistor programme is found.
Moreover, in the case of real Riemannian four-manifolds, the
local theory of primary and secondary potentials for gravitino
fields, recently proposed by Penrose, has been applied to
Ricci-flat backgrounds with boundary.
The geometric interpretation
of the differential equations obeyed by such secondary potentials
is related to the analysis of integrability conditions in the
theory of massless fields, and
might lead to a better understanding of twistor geometry.
Thus, new tools are available in complex general relativity
and in classical field theory in real Riemannian backgrounds.}

\leftskip=0mm
\rightskip=0mm

\baselineskip=5truemm

\def\cstok#1{\leavevmode\thinspace\hbox{\vrule\vtop{\vbox{\hrule\kern1pt
\hbox{\vphantom{\tt/}\thinspace{\tt#1}\thinspace}}
\kern1pt\hrule}\vrule}\thinspace}
\vskip 1cm
\noindent
\leftline {\bf 1. Introduction}
\vskip 1cm
\noindent
Within the framework of relativistic theories of gravitation,
the main aim of the Penrose twistor programme is to provide a
purely geometric tool for solving the Einstein equations within
a holomorphic, conformally invariant formalism.$^{1,2}$ For this
purpose, the Lorentzian space-time of Einstein's general relativity
is replaced by a four-complex-dimensional complex-Riemannian
manifold with holomorphic metric, holomorphic connection and
holomorphic curvature tensor. Such a programme succeeds in studying
anti-self-dual space-times with or without a cosmological constant,
as well as in the analysis of the massless free-field equations of
classical field theory. However, the main problem is that, unless
half of the conformal curvature vanishes (i.e. the self-dual Weyl
spinor ${\widetilde \psi}_{A'B'C'D'}$), no twistor-space
description of general relativity can be given. More recently, in
the work appearing in Refs. 3--4,
Lorentzian and complex general relativity
have been studied within a multisymplectic framework.
The multisymplectic analysis is very helpful to study the gravitational
field, viewed as a constrained system, in a manifestly covariant
way. This means that no restrictive assumption on the space-time
topology is made, and the invariance group remains the full
diffeomorphism group of four-dimensional space-time.$^{3}$
Note that the analysis in Refs. 3--4 differs substantially from the
recent approaches to canonical gravity, where one takes instead
complex tetrads on a four-real-dimensional Lorentzian manifold.

The resulting
description of complex general relativity is outlined in the
following section, following Ref. 4, while new perspectives for
twistor theory in real Riemannian four-manifolds with boundary
are described in Secs. 3 and 4. These sections deal with
a recent analysis of Rarita-Schwinger potentials in curved
backgrounds, and are relevant both for twistor theory and
for the understanding of consistent supergravity theories.
\vskip 1cm
\leftline {\bf 2. Multimomenta for Complex General Relativity}
\vskip 1cm
\noindent
In our geometric framework
one starts from a one-jet bundle $J^{1}$ which, in local coordinates,
is described by a holomorphic coordinate system, with holomorphic
tetrad $e_{a}^{\; \; {\hat a}}$, holomorphic connection one-form
$\omega_{a}^{\; \; {\hat b}{\hat c}}$, multivelocities
corresponding to $e_{a}^{\; \; {\hat a}}$ and multivelocities
corresponding to $\omega_{a}^{\; \; {\hat b}{\hat c}}$, both of
holomorphic nature. The intrinsic form of the field equations,
which is a generalization of a mathematical structure already
existing in classical mechanics, leads to the complex vacuum
Einstein equations $R_{ab}=0$, and to a condition on the covariant
divergence of the multimomenta. Moreover, the
multimomentum map,$^{3}$ when evaluated on a section of $J^{1}$
and integrated on an arbitrary three-complex-dimensional
surface $\Sigma_{c}$, leads to the
holomorphic equations corresponding to the constraint equations
of the Lorentzian theory,
and reflects the invariance of complex general relativity under
all holomorphic coordinate transformations. The Hamiltonian
constraint is then replaced by a geometric structure
linear in the {\it holomorphic multimomenta}
providing two boundary terms in
these holomorphic equations can be
set to zero. For this purpose, one of the following three
conditions should hold:$^{4}$
\vskip 0.3cm
\noindent
(i) $\Sigma_{c}$ has no boundary;
\vskip 0.3cm
\noindent
(ii) the holomorphic multimomenta
${\tilde p}_{\; \; \; {\hat c}{\hat d}}^{ab}
\equiv ({\rm det} \; e)
\Bigr(e_{\; \; {\hat c}}^{a} \; e_{\; \; {\hat d}}^{b}
-e_{\; \; {\hat d}}^{a} \; e_{\; \; {\hat c}}^{b}\Bigr)$,
defined as the derivatives of the Lagrangian with respect
to the multivelocities corresponding to the connection
one-form,
vanish at $\partial \Sigma_{c}$ (and hence everywhere on
$\Sigma_{c}$, by virtue of a well-known theorem in complex
analysis);
\vskip 0.3cm
\noindent
(iii) denoting by ${\lambda}^{{\hat a}{\hat b}}$ the elements of
the algebra $o(4,C)$ corresponding to
the gauge group, their spinorial version
$\lambda^{AA'BB'}$, and the spinor form of the connection
one-form, vanish at $\partial \Sigma_{c}$. On using
two-component spinor notation, these boundary conditions take
the form
$$
\Lambda^{(CD)}=0 \; \; \; \;
{\widetilde \Lambda}^{(C'D')}=0
\; \; \; \; \rm {at} \; \partial \Sigma_{c}
\; \; \; \; ,
\eqno (2.1)
$$
$$
\Omega_{f}^{(CD)}=0 \; \; \; \;
{\widetilde \Omega}_{f}^{(C'D')}=0
\; \; \; \; \rm {at} \; \partial \Sigma_{c}
\; \; \; \; ,
\eqno (2.2)
$$
where $\lambda^{CC'DD'}=\Lambda^{(CD)} \; \epsilon^{C'D'}
+{\widetilde \Lambda}^{(C'D')} \; \epsilon^{CD}$,
and $\omega_{f}^{CC'DD'}=\Omega_{f}^{(CD)} \; \epsilon^{C'D'}
+{\widetilde \Omega}_{f}^{(C'D')} \; \epsilon^{CD}$.
The boundary conditions (2.2) may be replaced by the condition
$u^{AA'}=0$ at $\partial \Sigma_{c}$, where $u$ is a vector
field describing holomorphic coordinate transformations on
complex space-time. In other words the work in Ref. 4 shows that,
if $\Sigma_{c}$ has a boundary (which cannot be ruled out),
the holomorphic multimomenta should vanish on the whole of
$\Sigma_{c}$, to avoid having restrictions at $\Sigma_{c}$
on the spinor fields expressing the invariance of the theory
under all holomorphic coordinate transformations.

Interestingly, to ensure that the holomorphic multimomenta
vanish at $\partial \Sigma_{c}$, one obtains conditions
which admit, as a subset, the totally null two-complex-dimensional
surfaces known as $\alpha$-surfaces and $\beta$-surfaces. The
integrability condition for $\alpha$-surfaces is the vanishing
of the self-dual Weyl spinor, and hence the multisymplectic
formalism enables one to recover the anti-self-dual space-time
relevant for twistor theory. However, if $\partial \Sigma_{c}$
is not totally null, the resulting theory does not correspond to
twistor theory, and one has to study the topology and the
geometry of the space of two-complex-dimensional surfaces
$\partial \Sigma_{c}$ in the generic case. Moreover, one has to
solve a set of equations which are now linear in the
holomorphic multimomenta, both in classical and in quantum
gravity (as we said before,
they correspond to the constraint equations of the
Lorentzian theory). In the classical holomorphic theory, such
equations take the form$^{4}$
$$
\int_{\Sigma_{c}}\lambda^{{\hat c}{\hat d}}
\Bigr(D_{a}{\tilde p}^{ab}\Bigr)_{{\hat c}{\hat d}}
\; d^{3}x_{b}=0
\; \; \; \; ,
\eqno (2.3)
$$
$$
\int_{\Sigma_{c}}{\rm Tr} \biggr[{\tilde p}^{af} \Omega_{ad}
-{1\over 2}{\tilde p}^{ab}\Omega_{ab} \; \delta_{d}^{f}\biggr]
u^{d} \; d^{3}x_{f}=0
\; \; \; \; .
\eqno (2.4)
$$
With our notation, $\Omega_{ab}^{\; \; \; {\hat c}{\hat d}}$
is the holomorphic curvature of the holomorphic connection
one-form $\omega_{a}^{\; \; {\hat c}{\hat d}}$. Moreover,
$D$ is a connection which annihilates the internal-space
metric $\eta_{{\hat a}{\hat b}}$.$^{4}$ It should be emphasized
that these equations, resulting from the holomorphic version
of the multimomentum map, cannot be related to a Cauchy
problem as in the Lorentzian theory. Hence their interpretation,
as well as the proposal to set to zero the integral of the
holomorphic multimomentum map on an arbitrary
three-complex-dimensional surface,$^{4}$ deserve further thinking.

Interestingly, the analysis in Ref. 4 shows that a deep link exists
between complex space-times which are not anti-self-dual and
two-complex-dimensional surfaces which are not totally null.
In other words, on going beyond twistor theory, the analysis
of two-complex-dimensional surfaces still plays a key role.
However, we do not yet know how to write and solve the
operator version of the Eqs.
(2.3)--(2.4), and the question is,
of course, crucial for the whole multisymplectic programme.
\vskip 10cm
\leftline {\bf 3. Local Theory of Spin-${3\over 2}$ Potentials}
\vskip 1cm
\noindent
Penrose has recently proposed a new definition of twistors
as charges for massless spin-${3\over 2}$ fields in
Ricci-flat Riemannian manifolds.$^{2,5}$ We now show that
the Penrose formalism can be applied to Ricci-flat
backgrounds with boundary, which are relevant for one-loop
quantum cosmology and for the analysis of consistent
supergravity theories.

The basic ideas
of the local theory of Rarita-Schwinger potentials are as
follows.$^{2}$ The independent spinor-valued one-forms
$\Bigr(\psi_{a}^{A},{\widetilde \psi}_{a}^{A'}\Bigr)$
occurring in the action functional of supergravity are obtained
from the tetrad and from some spinor fields as
$$
\psi_{a}^{A}=\Gamma_{\; \; \; \; \; \; B}^{C'A}
\; e_{\; \; C'a}^{B}
\; \; \; \; ,
\eqno (3.1)
$$
$$
{\widetilde \psi}_{a}^{A'}=\gamma_{\; \; \; \; \; B'}^{CA'}
\; e_{C \; \; \; a}^{\; \; B'}
\; \; \; \; .
\eqno (3.2)
$$
By virtue of the spinor Ricci identities and of the local
equations
$$
\Gamma_{B \; \; \; B'}^{\; \; A}
=\nabla_{BB'} \; \alpha^{A}
\; \; \; \; ,
\eqno (3.3)
$$
$$
\gamma_{B' \; \; \; B}^{\; \; \; A'}
=\nabla_{BB'} \; {\widetilde \alpha}^{A'}
\; \; \; \; ,
\eqno (3.4)
$$
the primary potentials $\gamma$ and $\Gamma$ obey the
differential equations
$$
\epsilon^{B'C'} \; \nabla_{A(A'} \;
\gamma_{\; \; B')C'}^{A}
=-3\Lambda \; {\widetilde \alpha}_{A'}
\; \; \; \; ,
\eqno (3.5)
$$
$$
\nabla^{B'(B} \; \gamma_{\; \; \; B'C'}^{A)}
=\Phi_{\; \; \; \; \; \; \; C'}^{ABL'}
\; {\widetilde \alpha}_{L'}
\; \; \; \; ,
\eqno (3.6)
$$
$$
\epsilon^{BC} \; \nabla_{A'(A} \;
\Gamma_{\; \; \; B)C}^{A'}=-3\Lambda \; \alpha_{A}
\; \; \; \; ,
\eqno (3.7)
$$
$$
\nabla^{B(B'} \; \Gamma_{\; \; \; \; BC}^{A')}
={\widetilde \Phi}_{\; \; \; \; \; \; \; \; \; C}^{A'B'L}
\; \alpha_{L}
\; \; \; \; ,
\eqno (3.8)
$$
where the spinor fields $\Phi$ and $\widetilde \Phi$
describe the trace-free Ricci spinor.
Note that $\Bigr(\alpha_{A},{\widetilde \alpha}_{A'}\Bigr)$
are a pair of independent and anticommuting spinor fields.
The primary potentials are subject to the gauge transformations
$$
{\widehat \gamma}_{\; \; B'C'}^{A} \equiv
\gamma_{\; \; B'C'}^{A}+\nabla_{\; \; B'}^{A}
\; \lambda_{C'} \; \; \; \; ,
\eqno (3.9)
$$
$$
{\widehat \Gamma}_{\; \; \; BC}^{A'} \equiv
\Gamma_{\; \; \; BC}^{A'}
+\nabla_{\; \; \; B}^{A'} \; \nu_{C} \; \; \; \; ,
\eqno (3.10)
$$
where the spinor fields $\nu_{B}$ and $\lambda_{B'}$ are
{\it freely specifiable}. Thus, the gauge transformations
(3.9)--(3.10) are compatible with the field equations
if and only if the scalar curvature and the trace-free
part of the Ricci tensor vanish, which implies that the
background geometry has to be Ricci-flat.
A set of secondary potentials is now introduced locally by
requiring that (cf. Eqs. (3.3)--(3.4))
$$
\gamma_{A'B'}^{\; \; \; \; \; \; \; C}
\equiv \nabla_{BB'} \; \rho_{A'}^{\; \; \; CB}
\; \; \; \; ,
\eqno (3.11)
$$
$$
\Gamma_{AB}^{\; \; \; \; \; C'} \equiv
\nabla_{BB'} \; \theta_{A}^{\; \; C'B'}
\; \; \; \; .
\eqno (3.12)
$$
If one now inserts Eqs. (3.11)--(3.12) into Eqs. (3.5)--(3.8) one
finds that, providing the following conditions hold (see appendix):
$$
\nabla^{B'(F} \; \rho_{B'}^{\; \; \; A)L}=0
\; \; \; \; ,
\eqno (3.13)
$$
$$
\nabla^{B(F'} \; \theta_{B}^{\; \; A')L'}=0
\; \; \; \; ,
\eqno (3.14)
$$
the secondary potentials obey the equations$^{2}$
$$
\psi^{ABLM} \; \rho_{(LM)C'}=0
\; \; \; \; ,
\eqno (3.15)
$$
$$
{\widetilde \psi}^{A'B'L'M'} \; \theta_{(L'M')C}=0
\; \; \; \; .
\eqno (3.16)
$$
Interestingly, no further restriction
on the curvature of the background
is obtained providing the symmetric parts of the secondary
potentials vanish, i.e.
$\rho_{\; \; \; \; \; \; \; C'}^{(AB)}=0$,
$\theta_{\; \; \; \; \; \; \; \; \; \; C}^{(A'B')}=0$,
and providing $\alpha_{A},
{\widetilde \alpha}_{A'}$ obey the Weyl equations
$$
\nabla^{AA'} \; \alpha_{A}=0
\; \; \; \; , \; \; \; \;
\nabla^{AA'} \; {\widetilde \alpha}_{A'}=0
\; \; \; \; .
\eqno (3.17)
$$
This implies that the secondary potentials take the form
$$
\rho_{A'}^{\; \; \; CB}=\epsilon^{CB} \;
{\widetilde \alpha}_{A'}
\; \; \; \; ,
\eqno (3.18)
$$
$$
\theta_{A}^{\; \; C'B'}=\epsilon^{C'B'} \;
\alpha_{A}
\; \; \; \; .
\eqno (3.19)
$$
However, if one wants to consider secondary potentials in their
complete form, Eqs. (3.15) and (3.16) imply that only
flat Euclidean four-space can be studied, since
otherwise one would obtain secondary potentials which depend
explicitly on the curvature of the background, and this is
inconsistent.

Moreover, the gauge freedom is restricted by the presence of
boundaries, since the boundary conditions should be preserved
by the gauge transformations (3.9)--(3.10). On imposing the
boundary conditions motivated by local supersymmetry$^{2}$
$$
\sqrt{2} \; {_{e}n_{A}^{\; \; A'}} \; \psi_{i}^{A}
= \pm {\widetilde \psi}_{i}^{A'}
\; \; \; \; {\rm at} \; \; \; \; \partial M \; \; \; \; ,
\eqno (3.20)
$$
one thus finds the following restrictions on the spinor
fields $\nu_{B}$ and $\lambda_{B'}$:
$$
\sqrt{2} \; {_{e}n_{A}^{\; \; A'}}
\Bigr(\nabla^{AC'} \; \nu^{B}\Bigr)e_{BC'i}
=\pm \Bigr(\nabla^{CA'} \; \lambda^{B'}\Bigr)
e_{CB'i} \; \; \; \; {\rm at} \; \; \; \;
\partial M \; \; \; \; .
\eqno (3.21)
$$
\vskip 1cm
\leftline {\bf 4. Open Problems}
\vskip 1cm
\noindent
Although we have presented only a very brief outline of the
local theory of spin-${3\over 2}$ potentials,
many interesting questions arise already at this stage:
\vskip 0.3cm
\noindent
(i) Can one find a complete two-spinor description of massive
spin-${3\over 2}$ potentials in Einstein backgrounds with
non-vanishing cosmological constant ? For this purpose, one
has to introduce a new covariant derivative, which differs
from the original one by a term proportional to the curved-space
$\gamma$-matrices. The two-component spinor formulation of the
resulting set of equations for spin-${3\over 2}$ potentials is
highly non-trivial, and is being investigated by myself and
G. Pollifrone.
\vskip 0.3cm
\noindent
(ii) Can one relate Eqs. (3.13)--(3.14) to the theory of integrability
conditions relevant for massless fields in curved backgrounds
(see appendix) ?
What happens when such equations do not hold ?
\vskip 0.3cm
\noindent
(iii) Is there an underlying global theory ? In the affirmative
case, what are the key features of the global theory ?
\vskip 0.3cm
\noindent
(iv) Can one define twistors as charges$^{2}$ for spin ${3\over 2}$
in Ricci-flat backgrounds with boundary ?
\vskip 0.3cm
\noindent
(v) Can one reconstruct the Riemannian four-geometry from the
twistor space, or from whatever is going to replace twistor
space ?

The solution of these problems might
improve our understanding of
the geometric properties relevant for classical
and quantum gravity. When combined with the ideas described
in the first part of this paper, these investigations seem to
suggest that a new synthesis is in sight in relativistic theories
of gravitation.
\vskip 1cm
\leftline {\bf Appendix}
\vskip 1cm
\noindent
The local theory of Rarita-Schwinger potentials leads naturally
to the consideration of Eq. (3.13) (and similarly for Eq. (3.14)),
since the insertion of Eq. (3.11) into Eq. (3.5) yields
in the Ricci-flat case
$$
\epsilon_{FL} \; \nabla_{AA'} \;
\nabla^{B'(F} \; \rho_{B'}^{\; \; \; A)L}
+{1\over 2}\nabla_{\; \; A'}^{A} \;
\nabla^{B'M} \; \rho_{B'(AM)}
+\cstok{\ }_{AM} \; \rho_{A'}^{\; \; \; (AM)}
+{3\over 8} \cstok{\ } \rho_{A'}=0
\; \; \; \; .
\eqno (A.1)
$$
Thus, if Eq. (3.13) holds, Eq. (A.1) reduces to
an identity by virtue of Ricci-flatness.

In the original approach by Penrose,$^{5}$ one describes
Rarita-Schwinger potentials in flat space-time in terms of
a rank-3 vector bundle with local coordinates
$\Bigr(\eta_{A},\zeta \Bigr)$, and an operator $\Omega_{AA'}$
whose action is defined by
$$
\Omega_{AA'}(\eta_{B},\zeta) \equiv
\Bigr({\cal D}_{AA'} \; \eta_{B}, {\cal D}_{AA'}\zeta
-\eta^{C} \; \rho_{A'AC} \Bigr)
\; \; \; \; ,
\eqno (A.2)
$$
where $\cal D$ is the flat Levi-Civita connection of Minkowski
space-time. The gauge transformations are then
$$
\Bigr({\widehat \eta}_{B},{\widehat \zeta}\Bigr) \equiv
\Bigr(\eta_{B}, \zeta +\eta_{A}\xi^{A} \Bigr)
\; \; \; \; ,
\eqno (A.3)
$$
$$
{\widehat \rho}_{A'AB} \equiv \rho_{A'AB}
+{\cal D}_{AA'} \; \xi_{B}
\; \; \; \; .
\eqno (A.4)
$$
For the operator $\Omega_{AA'}$ defined in
Eq. (A.2), the integrability
condition on $\beta$-planes turns out to be
$$
{\cal D}^{A'(A} \; \rho_{A'}^{\; \; \; B)C}=0
\; \; \; \; .
\eqno (A.5)
$$
It now remains to be seen whether, in a curved background,
an operator can be defined (cf. Eq. (A.2)) whose integrability
condition on $\beta$-surfaces is indeed given by Eq. (3.13)
(cf. Eq. (A.5)).
\vskip 1cm
\leftline {\bf Acknowledgments}
\vskip 1cm
\noindent
I am much indebted to Gabriele Gionti, Alexander Kamenshchik,
Giuseppe Marmo, Igor Mishakov, Giuseppe
Pollifrone and Cosimo Stornaiolo for scientific collaboration
on the topics described in this paper, and to Pietro
Santorelli for advice on the use of Latex.
Financial support by
the European Union under the Human Capital and Mobility Programme
is gratefully acknowledged.
\vskip 1cm
\leftline {\bf References}
\vskip 1cm
\begin{description}
\item [1.]
R. Penrose and W. Rindler, {\it Spinors and Space-Time II:
Spinor and Twistor Methods in Space-Time Geometry}
(Cambridge University Press, Cambridge, 1986).
\item [2.]
G. Esposito, {\it Complex General Relativity},
Fundamental Theories of Physics, Volume 69
(Kluwer, Dordrecht, 1995).
\item [3.]
G. Esposito, G. Gionti and C. Stornaiolo, {\it Spacetime Covariant
Form of Ashtekar's Constraints} (DSF preprint 95/7, to appear in
{\it Nuovo Cimento B}).
\item [4.]
G. Esposito and C. Stornaiolo, {\it Boundary Terms in Complex
General Relativity} (DSF preprint 94/58, to appear in
{\it Class. Quantum Grav.}).
\item [5.]
R. Penrose, in {\it Twistor Theory}, ed. S. Huggett
(Dekker, New York, 1994).
\end{description}

\end{document}